\documentclass[12pt,onecolumn,a4paper]{article}
\usepackage{sw20aip}

\input tcilatex
\QQQ{Language}{
American English
}

\begin{document}

\title{Semiclassical Green functions in mixed spaces}
\author{Guangcan Yang \\
Laboratory of Optical Physics, Institute of Physics, \\
Chinese Academy of Sciences, \\
P. O. Box 603, Beijing, 100080, China.}
\maketitle

\begin{abstract}
A explicit formula on semiclassical Green functions in mixed position and
momentum spaces is given based on Maslov's multi-dimensional semiclassical
theory. The general formula includes both coordinate and momentum
representations of Green functions as two special cases of the form.

\end{abstract}

Semiclassical methods are as old as quantum mechanics and an enormous amount
of work has been done in the field. Among them, the WKB method can be found
in any standard quantum mechanics textbooks\cite{WKB}. But it is usually
confined in one dimensional case. There has long been a need for the
development of a method that can be used to deal with multi-dimensional
problems. In last two decades, the Maslov and Fedoriuk's (MF) \textit{n-}%
dimensional semiclassical theory\cite{MF} has achieved a remarkable success
in atomic and molecular physics field\cite{Du}. The semiclassical method
emphasizes a suitable choice of representation space (usually a mixed
position-momentum space) to overcome the semiclassical diversity in
configuration or momentum space. Delos has shown how to construct a uniform
semiclassical wave function by MF's approach\cite{Delos}. The semiclassical
Green functions or propagators in configuration and momentum space has been
available for a long time\cite{Gutzwiller}. But they can not work in the
singular regions such as near caustics or foci of classical trajectories.
Semiclassical Green functions and trace formulas in connection with caustics
are analyzed in detail by Littlejohn and coworkers\cite{Littlejohn1}\cite
{Littlejohn2}. They emphasize geometrical properties of the semiclassical
trace formula and shows the intrinsic property of Maslov index. In this
paper we want to construct a semiclassical Green function explicitly in
mixed position-momentum spaces, which can be used locally to fix the
diversity of its counterpart in configuration or momentum space at any
boundaries between classically allowed and classically forbidden regions.
The Green function in mixed space is important to construct to a uniform
semiclassical Green function in customary configuration space. In fact, in
above mentioned singular regions we can obtain a uniform semiclassical Green
function in following steps: Firstly, we transform the primitive Green
function in configuration space into a suitable chosen mixed space. Then a
finite Green function in the mixed space is calculated. Finally, we
transform it back into configuration space, and sewing it with the primitive
Green function smoothly to construct a uniform semiclassical Green function.
In the following we will achieve the three steps explicitly.

For simplicity's sake we consider a single particle without spin in $n$%
-dimensional configuration space, and its Hamiltonian independent of time.
The propagator at constant energy or Green function $G\left( q^{\prime
},q^{\prime \prime },E\right) $ satisfies the inhomogeneous equation

\begin{equation}
\lbrack E-H(p^{\prime \prime },q^{\prime \prime })]G\left( q^{\prime
},q^{\prime \prime },E\right) =\delta \left( q^{\prime \prime }-q^{\prime
}\right)  \label{1}
\end{equation}
where $q^{\prime }$ is the initial position, $q^{\prime \prime }$ the final
position and $E$ is the total energy. The semiclassical approximation $%
\tilde{G}$ of $G$ is given by\cite{Gutzwiller}

\begin{equation}
\tilde{G}\left( q^{\prime \prime },q^{\prime },E\right) =\frac{2\pi }{\left(
2\pi i\hbar \right) ^{(n+1)/2}}\sum_{class.\;traj.}\left( \left| D_s\right|
\right) ^{1/2}\exp \left[ \frac i\hbar \left( S-\mu \frac \pi 2\right)
\right]  \label{2}
\end{equation}
where

\begin{equation}
S\left( q^{\prime \prime },q^{\prime },E\right) =\int_{q^{\prime
}}^{q^{\prime \prime }}pdq,  \label{3}
\end{equation}
is the classical action evaluated along the classical path which leads from $%
q^{\prime }$ to $q^{\prime \prime }$ at the given energy $H(p,q)=E$ and $\mu 
$ is the Maslov index, which is related with topological property of the ray
of trajectories. Its geometrical properties has been discussed in detail by
Littlejohn and coworkers\cite{Littlejohn1}\cite{Littlejohn2}. The summation
in Eq.(\ref{2}) is over all classical trajectories from $q^{\prime }$ to $%
q^{\prime \prime }$. The determinant $D_s$ is defined as

\begin{equation}
D_s=\left| 
\begin{array}{c}
\frac{\partial ^2S}{\partial q^{\prime }\partial q^{\prime \prime }} \\ 
\frac{\partial ^2S}{\partial E\partial q^{\prime \prime }}
\end{array}
\begin{array}{c}
\frac{\partial ^2S}{\partial q^{\prime }\partial E} \\ 
\frac{\partial ^2S}{\partial E^2}
\end{array}
\right| .  \label{4}
\end{equation}
which is a $n+1$ th determinant and can be considerably simplified as $n-1$%
th one if introducing a local coordinate system centered on the trajectory%
\cite{Gutzwiller}\cite{Littlejohn1}\cite{Littlejohn2}. Denote the
coordinates $\left( \mathbf{y},z\right) $ , with $z$ runs along the
trajectory and $n-1$ coordinates $\mathbf{y}$ transverse to the path in such
a way $y=0$ specifies the path. In the local coordinate,

\begin{equation}
D_s=\frac 1{\left| \dot{z}^{\prime }\dot{z}^{\prime \prime }\right| }\left| 
\frac{\partial ^2S}{\partial \mathbf{y}^{\prime }\partial \mathbf{y}^{\prime
\prime }}\right| .
\end{equation}
In fact, above restrictions on the local coordinate are not necessary and
have been shown in Refs.(6) and (7).

The semiclassical propagator in momentum space is given , in analogy of its
counterpart in configuration space, by

\begin{equation}
\tilde{F}\left( p^{\prime \prime },p^{\prime },E\right) =\frac{2\pi }{\left(
2\pi i\hbar \right) ^{(n+1)/2}}\sum_{class.\;traj.}\left( \left| D_T\right|
\right) ^{1/2}\exp \left[ \frac i\hbar \left( T-\nu \frac \pi 2\right)
\right] ,  \label{5}
\end{equation}
where

\begin{equation}
T\left( p^{\prime \prime },p^{\prime },E\right) =-\int_{p^{\prime
}}^{p^{\prime \prime }}qdp,  \label{6}
\end{equation}
is the classical action along the classical path from $p^{\prime }$ to $%
p^{\prime \prime }$ at the given energy $H(p,q)=E$ and $\nu $ is the Maslov
index calculated in momentum space. Usually, $\nu \neq \mu $ because of the
different topological properties of trajectories between configuration and
momentum spaces. The determinant $D_T$ is defined in an analogy way as

\begin{equation}
D_T=\left| 
\begin{array}{c}
\frac{\partial ^2T}{\partial p^{\prime }\partial p^{\prime \prime }} \\ 
\frac{\partial ^2T}{\partial E\partial p^{\prime \prime }}
\end{array}
\begin{array}{c}
\frac{\partial ^2T}{\partial p^{\prime }\partial E} \\ 
\frac{\partial ^2T}{\partial E^2}
\end{array}
\right| .  \label{7}
\end{equation}
Applying a similar local coordinate in momentum space, say, $p_z$ and $%
\mathbf{p}_y$ are along and transverse the trajectory respectively, we have
the stability matrix in momentum space

\begin{equation}
D_T=\frac 1{\left| \dot{p}_z^{\prime }\dot{p}_z^{\prime \prime }\right|
}\left| \frac{\partial ^2T}{\partial \mathbf{p}_y^{\prime }\partial \mathbf{p%
}_y^{\prime \prime }}\right|
\end{equation}

According quantum mechanics Green functions both in position and momentum
spaces must be finite. However, semiclassical approximation of them contain
divergences at any boundary between classically allowed and classically
forbidden regions, and at any caustics or foci of classical trajectories.
The reason causing such divergences is not the semiclassical theory itself,
but configuration or momentum representations are not suitable in these
singular regions. If we transform the problem to some mixed
position-momentum representation, such divergences can be repaired. It has
been shown that there ``almost always'' exists a representation in which the
semiclassical approximation does not divergent in these singular regions\cite
{MF}\cite{Delos}.

We choose mixed coordinates $\left( p_\alpha ,q_\beta \right) $, where $%
\alpha =1,\cdots ,k$ and $\beta =k+1,\cdots ,n,i.e,$ which are disjoints
set. So the set $p_\alpha q_\beta $ contains no canonical conjugate pairs.
The propagator or Green function in the mixed space can be acquired from the
counterpart in configuration space by partial Fourier transformation. The
propagator $F\left( p_\alpha ^{\prime \prime }q_\beta ^{\prime \prime
},p_\alpha ^{\prime }q_\beta ^{\prime }E\right) $ from $p_\alpha ^{\prime
}q_\beta ^{\prime }$ to $p_\alpha ^{\prime \prime }q_\beta ^{\prime \prime }$
while propagating with the energy $E$, is defined by

\begin{eqnarray}
F\left( p_\alpha ^{\prime \prime }q_\beta ^{\prime \prime },p_\alpha
^{\prime }q_\beta ^{\prime }E\right) &=&\left( 2\pi \hbar \right) ^{-k}\int
dq_\alpha ^{\prime \prime }\int dq_\alpha ^{\prime }G\left( q^{\prime \prime
}q^{\prime }E\right)  \label{8} \\
&&\times \exp \left[ i\left( p_\alpha ^{\prime }q_\alpha ^{\prime }-p_\alpha
^{\prime \prime }q_\alpha ^{\prime \prime }\right) /\hbar \right] . 
\nonumber
\end{eqnarray}
Its semiclassical approximation $\tilde{F}\left( p_\alpha ^{\prime \prime
}q_\beta ^{\prime \prime },p_\alpha ^{\prime }q_\beta ^{\prime }E\right) $
can be acquired by inserting $\tilde{G}\left( q^{\prime \prime }q^{\prime
}E\right) $ into Eq.(\ref{8}) and evaluating the integral by $k-$dimensional
stationary phase approximation(SPA). Firstly, we calculate the integral on $%
q_\alpha ^{\prime }$ , that is

\begin{equation}
\tilde{F}\left( q^{\prime \prime },p_\alpha ^{\prime }q_\beta ^{\prime
}E\right) \approx \left( 2\pi \hbar \right) ^{-k/2}\int dq_\alpha ^{\prime }%
\tilde{G}\left( q^{\prime \prime }q^{\prime }E\right) \exp \left[ i\left(
p_\alpha ^{\prime }q_\alpha ^{\prime }-p_\alpha ^{\prime \prime }q_\alpha
^{\prime \prime }\right) /\hbar \right] .  \label{9}
\end{equation}
where $\tilde{F}\left( q^{\prime \prime },p_\alpha ^{\prime }q_\beta
^{\prime }E\right) $ represents the integral after SPA. Inserting Eq.(\ref{2}%
) into the expression, the exponent becomes

\begin{equation}
\tilde{S}=S(q^{\prime \prime }q^{\prime }E)+p_\alpha ^{\prime }q_\alpha
^{\prime }-p_\alpha ^{\prime \prime }q_\alpha ^{\prime \prime }  \label{10}
\end{equation}
apart from the factor $i/\hbar $. It is noted that we adopt a single $\tilde{%
S}$ to denote the exponent before and after SPA. For convenience, Maslov
index and the summation over classical orbits have been dropped temporarily.
The SP points $q_{\alpha 0}^{^{\prime }}$ satisfy

\begin{equation}
\left. \frac{\partial S}{\partial q_\alpha ^{\prime }}\right| _{q_{\alpha
0}^{\prime }}=-p_\alpha ^{\prime },\;\;\;\alpha =1,\cdots ,k.  \label{11}
\end{equation}
After SPA, the preexponential factor can be written as

\begin{eqnarray}
&&\frac{2\pi }{\left( 2\pi i\hbar \right) ^{\left( 2n+1\right) /2}}\cdot 
\frac{\left( 2\pi i\hbar \right) ^{k/2}}{\left( 2\pi \hbar \right) ^{k/2}}%
\left( D_s\right) ^{1/2}\left( \frac{\partial ^2\tilde{S}}{\partial
^2q_\alpha ^{\prime }}\right) ^{-1/2}  \nonumber  \label{12} \\
&=&\frac{2\pi i^{k/2}}{\left( 2\pi i\hbar \right) ^{\left( 2n+1\right) /2}}%
\left( D_s\right) ^{1/2}\left( -\frac{\partial q_\alpha ^{\prime }}{\partial
p_\alpha ^{\prime }}\right) ^{1/2}.  \label{12}
\end{eqnarray}
To simplify the expression further, we rewrite $D_s$ as

\begin{equation}
D_s=\left| 
\begin{array}{c}
\frac{\partial ^2S}{\partial q_\alpha ^{\prime }\partial q_\alpha ^{\prime
\prime }} \\ 
\frac{\partial ^2S}{\partial q_\beta ^{\prime }\partial q_\alpha ^{\prime
\prime }} \\ 
\frac{\partial ^2S}{\partial E\partial q_\alpha ^{\prime \prime }}
\end{array}
\begin{array}{c}
\frac{\partial ^2S}{\partial q_\alpha ^{\prime }\partial q_\beta ^{\prime
\prime }} \\ 
\frac{\partial ^2S}{\partial q_\beta ^{\prime }\partial q_\beta ^{\prime
\prime }} \\ 
\frac{\partial ^2S}{\partial E\partial q_\beta ^{\prime \prime }}
\end{array}
\begin{array}{c}
\frac{\partial ^2S}{\partial q_\alpha ^{\prime }\partial E} \\ 
\frac{\partial ^2S}{\partial q_\beta ^{\prime }\partial E} \\ 
\frac{\partial ^2S}{\partial E^2}
\end{array}
\right| ,  \label{13}
\end{equation}
and the second derivatives of $S$ now should be replaced the derivatives of $%
\tilde{S}$. To achieve the replacement, we can use the following relations

\begin{eqnarray}
\frac{\partial S}{\partial q_\alpha ^{\prime }} &=&-p_\alpha ^{\prime }, 
\nonumber \\
\frac{\partial \tilde{S}}{\partial q_\alpha ^{\prime \prime }} &=&\frac{%
\partial S}{\partial q_\alpha ^{\prime \prime }},  \label{14} \\
\frac{\partial \tilde{S}}{\partial E} &=&\frac{\partial S}{\partial E}. 
\nonumber
\end{eqnarray}

Inserting Eqs.(\ref{13}) and (\ref{14}), Eq.(\ref{12}) becomes

\begin{equation}
\frac{2\pi i^{k/2}}{\left( 2\pi i\hbar \right) ^{\left( 2n+1\right) /2}}%
\left| 
\begin{array}{c}
\frac{\partial ^2\tilde{S}}{\partial p_\alpha ^{\prime }\partial q_\alpha
^{\prime \prime }} \\ 
\frac{\partial ^2S}{\partial q_\beta ^{\prime }\partial q_\alpha ^{\prime
\prime }} \\ 
\frac{\partial ^2S}{\partial E\partial q_\alpha ^{\prime \prime }}
\end{array}
\begin{array}{c}
\frac{\partial ^2\tilde{S}}{\partial p_\alpha ^{\prime }\partial q_\beta
^{\prime \prime }} \\ 
\frac{\partial ^2S}{\partial q_\beta ^{\prime }\partial q_\beta ^{\prime
\prime }} \\ 
\frac{\partial ^2S}{\partial E\partial q_\beta ^{\prime \prime }}
\end{array}
\begin{array}{c}
\frac{\partial ^2\tilde{S}}{\partial p_\alpha ^{\prime }\partial E} \\ 
\frac{\partial ^2S}{\partial q_\beta ^{\prime }\partial E} \\ 
\frac{\partial ^2S}{\partial E^2}
\end{array}
\right| ^{1/2}\equiv \frac{2\pi i^{k/2}}{\left( 2\pi i\hbar \right) ^{\left(
2n+1\right) /2}}\left( D_{\tilde{s}}^{\prime }\right) ^{1/2}  \label{15}
\end{equation}
It is noted that in the first line $S$ is replaced by $\tilde{S}$ . After
SPA, the integral in Eq.(\ref{9}) becomes

\begin{equation}
\tilde{F}\left( q^{\prime \prime },p_\alpha ^{\prime }q_\beta ^{\prime
}E\right) =\frac{2\pi i^{k/2}}{\left( 2\pi i\hbar \right) ^{\left(
2n+1\right) /2}}\left( D_{\tilde{s}}^{\prime }\right) ^{1/2}\exp \left(
\frac i\hbar \tilde{S}\right) ,  \label{16}
\end{equation}
where

\begin{equation}
\tilde{S}=S(q^{\prime \prime }q_{\alpha 0}^{\prime }q_\beta ^{\prime
}E)+p_\alpha ^{\prime }q_{\alpha 0}^{\prime }-p_\alpha ^{\prime \prime
}q_\alpha ^{\prime \prime }.
\end{equation}
is the value of Eq.(\ref{10}) at the first SP point. Secondly, we again
approximate the integral over $q_\alpha ^{\prime \prime }$ by SPA to get the
mixed Green function

\begin{equation}
\tilde{F}\left( p_\alpha ^{\prime \prime }q_\beta ^{\prime \prime },p_\alpha
^{\prime }q_\beta ^{\prime }E\right) \approx \left( 2\pi \hbar \right)
^{-k/2}\int dq_\alpha ^{\prime \prime }\tilde{F}\left( q^{\prime \prime
},p_\alpha ^{\prime }q_\beta ^{\prime }E\right) .
\end{equation}
Again $\tilde{F}\left( p_\alpha ^{\prime \prime }q_\beta ^{\prime \prime
},p_\alpha ^{\prime }q_\beta ^{\prime }E\right) $ denotes the integral after
SPA. The SP points now satisfy

\begin{equation}
\left. \frac{\partial S}{\partial q_\alpha ^{\prime \prime }}\right|
_{q_{\alpha 0}^{\prime \prime }}=-p_\alpha ^{\prime \prime },\;\;\;\alpha
=1,\cdots ,k.
\end{equation}
The relations in Eq.(\ref{14}) are replaced by

\begin{eqnarray}
\frac{\partial S}{\partial q_\alpha ^{\prime \prime }} &=&-p_\alpha ^{\prime
\prime }, \\[0.03in]
\frac{\partial \tilde{S}}{\partial q_\beta ^{\prime }} &=&\frac{\partial S}{%
\partial q_\beta ^{\prime }},  \nonumber \\
\frac{\partial \tilde{S}}{\partial q_\beta ^{\prime \prime }} &=&\frac{%
\partial S}{\partial q_\beta ^{\prime \prime }},  \nonumber \\
\frac{\partial \tilde{S}}{\partial E} &=&\frac{\partial S}{\partial E}, 
\nonumber
\end{eqnarray}
which leads to the replacement of the derivatives with respect to $q_\alpha
^{\prime \prime }$ in the first column of $D_{\tilde{S}}^{\prime }$ by the
ones to $p_\alpha ^{\prime \prime }$. Then the preexponential factor arrives
at

\begin{equation}
\frac{2\pi i^{k/2}}{\left( 2\pi i\hbar \right) ^{\left( 2n+1\right) /2}}%
\left| 
\begin{array}{c}
\frac{\partial ^2\tilde{S}}{\partial p_\alpha ^{\prime }\partial p_\alpha
^{\prime \prime }} \\ 
\frac{\partial ^2\tilde{S}}{\partial q_\beta ^{\prime }\partial p_\alpha
^{\prime \prime }} \\ 
\frac{\partial ^2\tilde{S}}{\partial E\partial p_\alpha ^{\prime \prime }}
\end{array}
\begin{array}{c}
\frac{\partial ^2\tilde{S}}{\partial p_\alpha ^{\prime }\partial q_\beta
^{\prime \prime }} \\ 
\frac{\partial ^2\tilde{S}}{\partial q_\beta ^{\prime }\partial q_\beta
^{\prime \prime }} \\ 
\frac{\partial ^2\tilde{S}}{\partial E\partial q_\beta ^{\prime \prime }}
\end{array}
\begin{array}{c}
\frac{\partial ^2\tilde{S}}{\partial p_\alpha ^{\prime }\partial E} \\ 
\frac{\partial ^2\tilde{S}}{\partial q_\beta ^{\prime }\partial E} \\ 
\frac{\partial ^2\tilde{S}}{\partial E^2}
\end{array}
\right| ^{1/2}\equiv \frac{2\pi }{\left( 2\pi i\hbar \right) ^{\left(
2n+1\right) /2}}\left( D_{\tilde{s}}\right) ^{1/2}
\end{equation}
It is noted that the factor $i^k$ after two integrations has been absorbed
in the final determinant $D_{\tilde{S}}$ . Therefore, the final Green
function in $p_\alpha q_\beta $ space is

\begin{equation}
\tilde{F}\left( p_\alpha ^{\prime \prime }q_\beta ^{\prime \prime },p_\alpha
^{\prime }q_\beta ^{\prime }E\right) =\frac{2\pi }{\left( 2\pi i\hbar
\right) ^{\left( 2n+1\right) /2}}\sum_{class.\;traj.}\left( \left| D_{\tilde{%
S}}\right| \right) ^{1/2}\exp \left[ \frac i\hbar \left( \tilde{S}-\tilde{\nu%
}\frac \pi 2\right) \right] ,  \label{final}
\end{equation}
where

\begin{equation}
\tilde{S}=S(q_{\alpha 0}^{\prime \prime }q_\beta ^{\prime \prime }q_{\alpha
0}^{\prime }q_\beta ^{\prime }E)+p_\alpha ^{\prime }q_{\alpha 0}^{\prime
}-p_\alpha ^{\prime \prime }q_{\alpha 0}^{\prime \prime }.  \label{action}
\end{equation}
is the value of Eq.(\ref{10}) at the second SP point though with the same
symbol. In Eq.(\ref{final}) the Maslov index $\tilde{\nu}$ has been added
and is related with the topological properties of classical trajectories in
the mixed space. The summation over classical trajectories has also been
attached. For convenience we have adopted the single symbol $\tilde{S}$ to
represent the phase factor before and after SPA. $\tilde{S}$ can be written
in a regular form 
\begin{equation}
\tilde{S}=\int_{q_\beta ^{\prime }}^{q_\beta ^{\prime \prime }}p_\beta
dq_\beta -\int_{p_\alpha ^{\prime }}^{p_\alpha ^{\prime \prime }}q_\alpha
dp_\alpha  \label{action1}
\end{equation}
which is the classical action evaluated the trajectory from $p_\alpha
^{\prime }q_\beta ^{\prime }$ to $p_\alpha ^{\prime \prime }q_\beta ^{\prime
\prime }$ at given energy $H\left( pq\right) =E$. It can be acquired by
differentiate Eq.(\ref{action}) with respect to the terminal coordinates.
Eq.(\ref{action}) can be rewritten as

\begin{equation}
\tilde{S}=\int_{q_{\alpha 0}^{\prime }}^{q_{\alpha 0}^{\prime \prime
}}p_\alpha dq_\alpha +\int_{q_\beta ^{\prime }}^{q_\beta ^{\prime \prime
}}p_\beta dq_\beta +p_\alpha ^{\prime }q_{\alpha 0}^{\prime }-p_\alpha
^{\prime \prime }q_{\alpha 0}^{\prime \prime }.
\end{equation}
which has contained the first term in Eq.(\ref{action1}). Differentiating it
with respect to $p_\alpha ^{\prime }$ , we have

\begin{equation}
\frac{\partial \tilde{S}}{\partial p_\alpha ^{\prime }}=-p_\alpha ^{\prime }%
\frac{\partial q_{a0}^{\prime }}{\partial p_\alpha ^{\prime }}+q_{\alpha
0}^{\prime }+p_\alpha ^{\prime }\frac{\partial q_{a0}^{\prime }}{\partial
p_\alpha ^{\prime }}=q_{\alpha 0}^{\prime }\left( p_\alpha ^{\prime \prime
}\right) .  \label{par1}
\end{equation}
Similarly, one finds that

\begin{equation}
\frac{\partial \tilde{S}}{\partial p_\alpha ^{\prime \prime }}=-q_{\alpha
0}^{\prime \prime }\left( p_\alpha ^{\prime \prime }\right) .  \label{par2}
\end{equation}
Eq.(\ref{par1}) and (\ref{par2}) immediately leads to (\ref{action1}).

The stability matrix $D_{\tilde{s}}$ can also be simplified as a $n-1$th
determinant depending on the choice of local coordinate centered on the
trajectory in the mixed space. If we choose a position coordinate $z$ along
with the trajectory in the mixed space, $\left( p_\alpha ,\mathbf{y}_\beta
\right) $ transverse the path, then

\begin{equation}
D_{\tilde{S}}=\frac 1{\left| \dot{z}^{\prime }\dot{z}^{\prime \prime
}\right| }\left| 
\begin{array}{c}
\frac{\partial ^2\tilde{S}}{\partial p_\alpha ^{\prime }\partial p_\alpha
^{\prime \prime }} \\ 
\frac{\partial ^2\tilde{S}}{\partial \mathbf{y}_\beta ^{\prime }\partial
p_\alpha ^{\prime \prime }}
\end{array}
\begin{array}{c}
\frac{\partial ^2\tilde{S}}{\partial p_\alpha ^{\prime }\partial \mathbf{y}%
_\beta ^{\prime \prime }} \\ 
\frac{\partial ^2\tilde{S}}{\partial \mathbf{y}_\beta ^{\prime }\partial
y_\beta ^{\prime \prime }}
\end{array}
\right| .
\end{equation}
Conversely, choosing a momentum coordinate $p_z$ along with the trajectory
and $\left( p_{y\alpha },q_\beta \right) $ transverse the path, we have

\[
D_{\tilde{S}}=\frac 1{\left| \dot{p}_z^{\prime }\dot{p}_z^{\prime \prime
}\right| }\left| 
\begin{array}{c}
\frac{\partial ^2\tilde{S}}{\partial p_{y\alpha }^{\prime }\partial
p_{y\alpha }^{\prime \prime }} \\ 
\frac{\partial ^2\tilde{S}}{\partial q_\beta ^{\prime }\partial p_{y\alpha
}^{\prime \prime }}
\end{array}
\begin{array}{c}
\frac{\partial ^2\tilde{S}}{\partial p_{y\alpha }^{\prime }\partial q_\beta
^{\prime \prime }} \\ 
\frac{\partial ^2\tilde{S}}{\partial q_\beta ^{\prime }\partial q_\beta
^{\prime \prime }}
\end{array}
\right| 
\]

Formula(\ref{final}) is a unified expression. If $\alpha =0,\beta =1\cdots
n, $ it will recover as Eq.(\ref{2}), the Green function in configuration
space. Conversely, if $\alpha =1\cdots n,$ $\beta =0,$ Eq.(\ref{5}), the
propagator in momentum space will immediately emerge.

For most semiclassical physical problems, it is convenient to work in
configuration space. Eq.(\ref{2}) usually works quite will in most regions.
But in some regions, such as near caustics and foci of classical
trajectories, it will be divergent. In these regions, we can `` almost
always''\cite{MF} \cite{Delos}choose a suitable set of mixed
position-momentum coordinates to calculate the mixed Green function, which
is not divergent. From the function, an accurate configuration-space Green
function can be constructed be inverse partial Fourier transformation,

\begin{eqnarray}
\tilde{G}\left( q^{\prime \prime }q^{\prime }E\right) &=&\left( 2\pi \hbar
\right) ^{-k}\int dp_\alpha ^{\prime \prime }\int dp_\alpha ^{\prime }\tilde{%
F}\left( p_\alpha ^{\prime \prime }q_\beta ^{\prime \prime },p_\alpha
^{\prime }q_\beta ^{\prime }E\right) \\
&&\times \exp \left[ i\left( -p_\alpha ^{\prime }q_\alpha ^{\prime
}+p_\alpha ^{\prime \prime }q_\alpha ^{\prime \prime }\right) /\hbar \right]
.  \nonumber
\end{eqnarray}
Then we can sew it with the primitive Green function smoothly to get a
uniform Green function.

The author thanks Professor M. L. Du for helpful discussions.

\end{document}